\documentclass[fleqn,usenatbib]{mnras}

\usepackage{newtxtext,newtxmath}

\usepackage[brazilian]{babel}
\usepackage[utf8]{inputenc}
\usepackage[T1]{fontenc}

\usepackage[T1]{fontenc}
\usepackage{ae,aecompl}

\usepackage{graphicx}	
\usepackage{amsmath}	
\usepackage{amssymb}	
\usepackage{gensymb}

\title[Eclipse timing variation of GK\,Vir]
 {Eclipse timing variation of GK\,Vir: evidence of a possible Jupiter-like planet in a circumbinary orbit.}
\author[L.~A.~Almeida et al.]
  {L.~A.~Almeida$^{1,2}$\thanks{E-mail: leonardoalmeida@uern.br}, E. S. Pereira$^{3}$, G.~M.~Borges$^{4}$, A.~Damineli$^{3}$, T.~A.~Michtchenko$^{3}$, \\ 
  \newauthor G.~M.~Viswanathan$^{2,5}$ 
   \\
   \\
  $^1$Departamento de F\'isica, Universidade do Estado do Rio Grande do Norte, Mossor\'o, RN 59610-210, Brazil \\
  $^2$Departamento de F\'isica, Universidade Federal do Rio Grande do Norte, Natal, RN 59072-970, Brazil \\
  $^3$Instituto de Astronomia, Geof\'isica e Ci\^encias Atmosf\'ericas, Rua do Mat\~ao 1226, S\~ao Paulo, SP 05508-090, Brazil\\
  $^4$Departamento de Ci\^encias Exatas e Tecnologia da Informa\c{c}\~ao, Universidade Federal Rural do Semi-\'Arido, Angicos, RN 59515-000,\\ Brazil \\
  $^5$National Institute of Science and Technology of Complex Systems, Universidade Federal do Rio Grande do Norte, Natal, RN~59072-970, Brazil
  } 

\date{Accepted 2020 July 16; Received 2020 March 7 in original form L.A. Almeida}

\pubyear{2020}
\hypersetup{draft}
\begin{document}
\label{firstpage}
\pagerange{\pageref{firstpage}--\pageref{lastpage}}
\maketitle

\begin{abstract}
Eclipse timing variation analysis has become a powerful method to discover planets around binary systems. We applied this technique to investigate the eclipse times of GK\,Vir. This system is a post-common envelope binary with an orbital period of 8.26~h. Here, we present 10 new eclipse times obtained between 2013 and 2020. We calculated the O-C diagram using a linear ephemeris and verified a clear orbital period variation (OPV) with a cyclic behavior. We investigated if this variation could be explained by the Applegate mechanism, the apsidal motion, or the light travel time (LTT) effect. We found that the Applegate mechanism would hardly explain the OPV with its current theoretical description. We obtained using different approaches that the apsidal motion is a less likely explanation than the LTT effect. We showed that the LTT effect with one circumbinary body is the most likely cause for the OPV, which was reinforced by the orbital stability of the third body. The LTT best solution provided an orbital period of $\sim$24~yr for the outer body. Under the assumption of coplanarity between the external body and the inner binary, we obtained a Jupiter-like planet around the GK\,Vir. In this scenario, the planet has one of the longest orbital periods, with a full observational baseline, discovered so far. However, as the observational baseline of GK\,Vir is smaller than twice the period found in the O-C diagram, the LTT solution must be taken as preliminary.
\end{abstract}


\section{Introduction}\label{introduction}

Orbital period variation (OPV) of post-common envelope binaries has become a powerful tool to search for circumbinary planets \citep[see, e.g.][]{2009AJ.137.3181L,Beuermann+2010,2010MNRAS.401L.34Q,Almeida+2013,Almeida+2019}, as well as to investigate intrinsic phenomena of the binary, e.g., the magnetic cycle of one active component, apsidal motion, mass transfer events, angular momentum loss via magnetic braking and gravitation wave emission \citep[see, e.g.,][]{Claret+2010,Parsons+2010,Schreiber+2010,Zorotovic+2013,Bours+2016,Almeida+2019,Burdge+2019}. While some of these phenomena induce a decrease or an increase in the orbital period of the binary, the magnetic cycle, the apsidal motion, and the gravitational influence of a third body produce cyclic and periodic variations. These three last effects have similar features in time-scales from months to decades and therefore difficult to be distinguished. Despite of some important clues to solve this issue, in both observational and theoretical sides, have been reported in the literature, e.g. \citet[][]{Applegate1992, Brinkworth2006, Parsons+2010, Parsons+2014, Bours+2016, Volschow+2016, Volschow+2018, Almeida+2019}, it is still an open question and more post-common envelope systems with long-baseline of eclipse time monitoring are needed.

GK\,Vir is a detached eclipsing binary system consisting of a white dwarf (WD -- primary) and a low-mass main-sequence star (secondary) with an orbital period of 8.27~h \citep{Green+1978}. This system was discovered in a survey for blue stars at high Galactic latitude \citep[][]{Green+1976} and it is located at $\sim$475~pc from us \citep{Gaia+2018}. \citet[][]{Fulbright+1993} collected spectroscopic data of GK\,Vir and together with photometric information reported by \citet{Green+1978} improved the physical property measurements of the system. With high-resolution spectroscopic data and optical and infrared photometric data, \citet[][]{Parsons+2012} characterised both components as well as the geometrical parameters of GK\,Vir. The authors derived the binary's inclination (i = 89$\fdg$5), masses and radii for both, primary and secondary components (M$_{\rm WD} = 0.564~\rm M_{\sun}$, M$_{\rm sec} = 0.116~\rm M_{\sun}$ and R$_{\rm WD} = 0.0170~\rm R_{\sun}$, R$_{\rm sec} = 0.155~\rm R_{\sun}$). Using evolutionary models, \citet[][]{Parsons+2012} obtained an effective temperature T$_{\rm eff} = 50000$~K and a carbon-oxygen core for the WD.

The eclipse times of GK\,Vir have a long history. \citet[][]{Green+1978} presented the first nine measurements from 1975 to 1978 with uncertainties varying from 1 to 10~s. \citet[][]{Parsons+2010} reported seven new eclipsing time measurements derived from the high-speed ULTRACAM photometric data collected between 2002 and 2007, which showed a slight decrease in the O-C~diagram. Another study by \citet[][]{Parsons+2012} listed out one new eclipse time measure obtained in 2010, which showed an increase in the O-C~diagram. The same trend in this diagram was found by \citet{Bours+2016} with 10 new measurements from 2012 to 2015. As pointed out by these last authors, GK\,Vir is the system with the second-largest observational baseline among the 58 eclipsing detached binaries, which are composed of a WD plus a main-sequence star or a brown dwarf. Also, in this sample, GK\,Vir has one of the smallest OPV amplitude. Therefore, this system is an important target to perform an OPV analysis.  

Here we present 10 new mid-eclipse times of GK\,Vir obtained between 2013 August and 2020 April. We combine these data with previous measurements from the literature and performed a new OPV analysis. In Section~\ref{observation}, we describe our data and the reduction procedure. The methodology used to obtain the mid-eclipse times, the procedure to examine the OPV, and their possible related physical effects are presented in Section~\ref{analysis}. In Section~\ref{discussion}, we discuss our results.

\section{Observations and data reduction}\label{observation}

The photometric data of GK\,Vir were collected in an observational program to search for OPVs in compact binaries. This project is carried out using the facilities of the \textit{Observat\'orio do Pico dos Dias} that is operated by the \textit{Laborat\'orio Nacional de Astrof\'isica} (LNA/MCTI) in Brazil. Photometric observations were performed with CCD cameras attached to the 0.6-m, and 1.6-m telescopes. The procedure to remove undesired effects from the CCD data includes typically obtaining 100 bias frames and $30$ dome flat-field images for each night of observations. The characteristics of the GK\,Vir photometric data are summarised in Table~\ref{table:1}. In this table, $n$ and $t_{\rm exp}$ are the number and the time of exposure, respectively. 

\begin{table}
\caption{Log of the photometric observations.} 
\footnotesize            
\label{table:1}      
\centering                          
\begin{tabular}{l c c c c}        
\hline\hline                 
Date~~~~~ & $n$ & \ t$_{\rm exp}$(s) & Telescope & Filter \\    
\hline                        
   2013 Aug 12 & 280  &  4 & 1.6-m & Unfiltered \\      
   2017 Apr 03 & 310  & 10 & 0.6-m & Unfiltered \\
   2017 Apr 05 & 120  & 10 & 0.6-m & Unfiltered \\
   2019 Apr 03 &  48  & 30 & 0.6-m & Unfiltered \\ 
   2019 Jun 02 & 216  & 10 & 0.6-m & Unfiltered \\   
   2019 Jul 14 & 377  & 10 & 1.6-m & Unfiltered \\ 
   2019 Jul 24 & 111  & 15 & 0.6-m & Unfiltered \\
   2020 Mar 04 & 289  & 15 & 0.6-m & Unfiltered \\
   2020 Mar 05 & 352  & 10 & 0.6-m & Unfiltered \\
   2020 Apr 01 & 151  & 20 & 0.6-m & Unfiltered \\
\hline                                   
\end{tabular}
\end{table}

The data reduction was done using the standard {\small IRAF}\footnote{IRAF is distributed by the National Optical Astronomy Observatory, which is operated by the Association of Universities for Research in Astronomy (AURA) under cooperative agreement with the National Science Foundation.} tasks. We have an automatic procedure to obtain the relative flux of GK\,Vir. The procedure consists of subtracting a master median bias image from each program image and dividing the result by a normalized flat-field. Then, differential photometry is used to obtain the relative flux between our target and a star of the field with constant flux. The extraction of the fluxes was done using aperture photometry as the GK\,Vir field is not crowded. Figure~\ref{mid-eclipse} shows the result of this procedure for the data collected on 2019 July 14. 

\section{Analysis and Results}\label{analysis}

\subsection{Eclipse fitting}
\label{eclipe_fit}
In order to obtain the mid-eclipse times of GK\,Vir, we performed a model fit for each observed event. The Wilson-Devinney code \citep[WDC;][]{1971ApJ.166.605W} was used to generate the synthetic light curves. As GK\,Vir is a detached binary, we used mode 2 of the WDC which is appropriate for such type of system. The luminosity of both components was computed assuming stellar atmosphere radiation. For the fitting procedure, we adopted as search intervals the range of the geometrical and physical parameters, e.g., mass ratio, inclination, radii, temperatures, and masses obtained by \citet{Parsons+2012} for GK\,Vir. 

\begin{figure}
\includegraphics[scale=0.46,angle=270,bb=80 100 420 340]{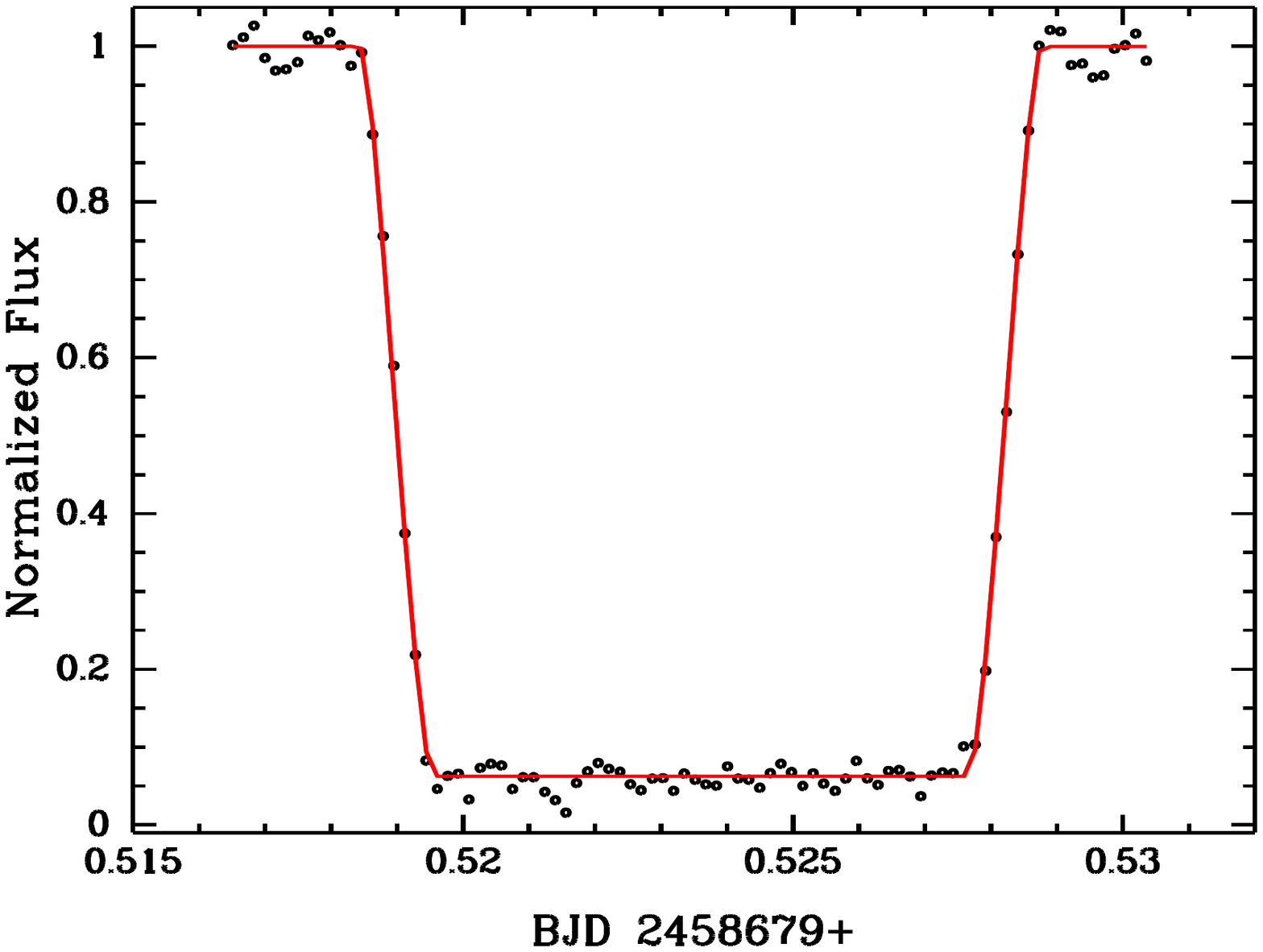}
	\caption{Light curve of the primary eclipse of GK\,Vir observed on 2019 July 14. The red line represents the best fit obtained using the procedure described in Section~\ref{eclipe_fit}.}
	\label{mid-eclipse}
\end{figure}

A procedure similar to that described in \citet{2012MNRAS.423.478A} was used to search for the mid-eclipse times of GK\,Vir. The light curves generated by the WDC were used as a `function'~to be optimized by the genetic algorithm \textsc{PIKAIA} \citep{char1995}. To measure the goodness of fit, we use the reduced chi-square defined as
\begin{equation}
\centering
\chi_{\rm red}^2= \frac{1}{(n-m)}\sum_1^n\left(\frac{O_j-C_j}{\sigma_j}\right)^2,
\end{equation}
where $O_j$ is the observed points, $C_j$ is the corresponding model, $\sigma_j$ is the uncertainties at each point, $n$ is the number of measurements, and $m$ is the number of fitted parameters. Figure~\ref{mid-eclipse} shows an eclipse of GK\,Vir with the best solution superimposed. To establish realistic uncertainties, we used the solution obtained by \textsc{PIKAIA} as input to a Markov chain Monte Carlo (MCMC) procedure \citep[][]{Foreman-Mackey2013} and examine the marginal posterior distribution of the probability of the parameters. The median of the distribution gives the time of mid-eclipse and the area corresponding to the $\sim$68.3\% in a normal distribution gives the corresponding standard uncertainty. The results are presented in Table~\ref{timing}.

\subsection{Orbital period variations}

\begin{figure*}
\begin{minipage}{175mm}
\resizebox{\hsize}{!}{\includegraphics[angle=0]{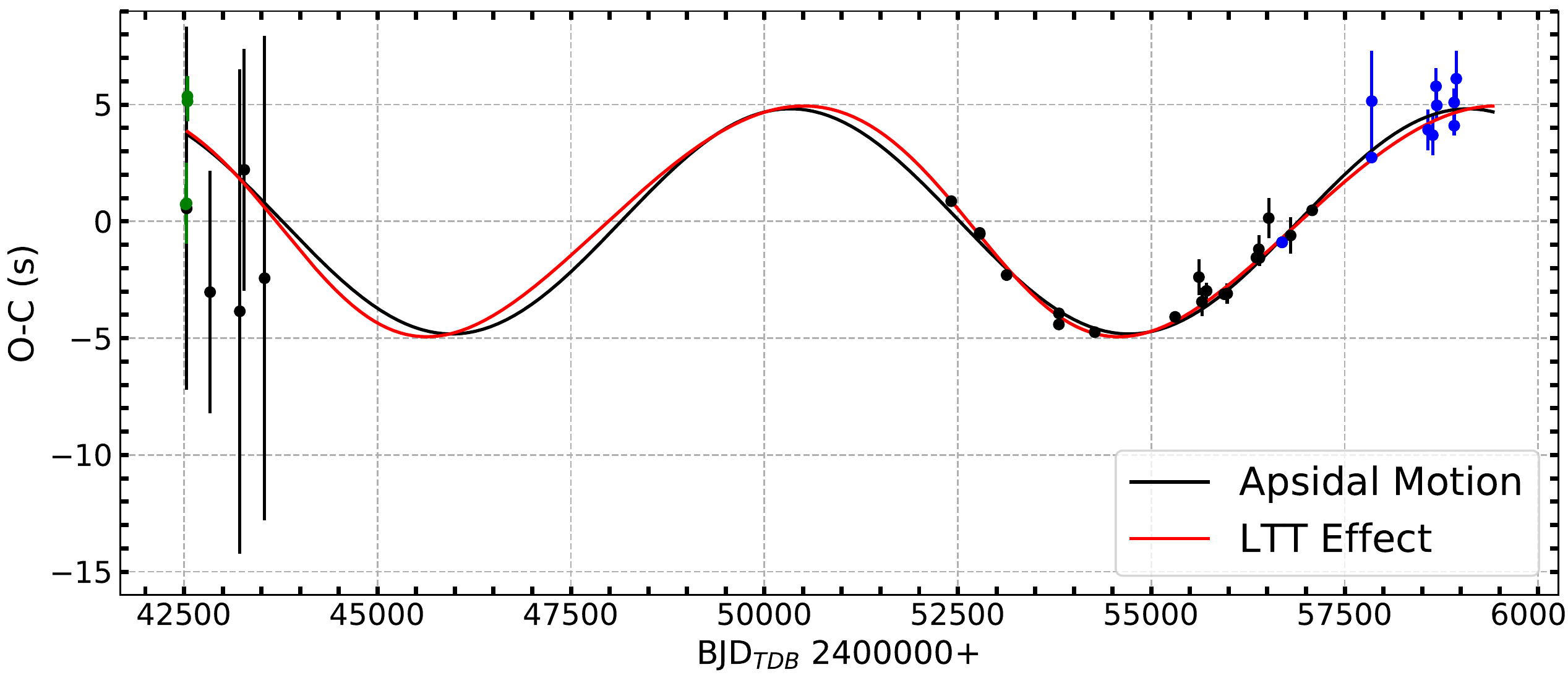}}
\caption{(O$-$C) diagram of the mid-eclipse times of GK\,Vir built with respect to the linear part of the ephemeris in Eqs.~\ref{ephem_ltt} and \ref{eqam}. Our measurements are presented with blue circles and the red and black lines represent the best fit including one LTT effect and the apsidal motion, respectively. The green and black points are measurements from \citet[][]{Green+1978}, \citet[][]{Parsons+2010,Parsons+2012}, and \citet[][]{Bours+2015}. Four measurements from \citet[][]{Green+1978} which have smaller error bars and are overlapped in this plot (see Table~\ref{timing}), were highlighted with green colour.}
\label{figo-c}
\end{minipage}{}
\end{figure*}

To analyse the OPVs of GK\,Vir, we collected the available mid-eclipse times in the literature and added the 10 new measurements. We first fit a linear ephemeris,
\begin{equation}
 T_{\rm min} = T_0 + E\times P_{\rm bin},
 \label{lin_efem}
\end{equation}
to all mid-eclipse times. In the last equation, $T_{\rm min}$ is the mid-eclipse time, $T_0$, $E$, and $P_{\rm bin}$ are the initial epoch, the cycle, and the orbital period of the binary, respectively. The residuals obtained from the fit show a cyclic variation (see Figure~\ref{figo-c}). This kind of variation can be explained by the light travel time (LTT) effect, the apsidal motion, or the Applegate mechanism. In the next sections, these three possible scenarios are discussed.

\begin{table}
\caption{Mid-eclipse times of GK\,Vir.} 
\label{timing}      
\centering \scriptsize                      
\begin{tabular}{l c c r}        
\hline\hline                 
Cycle &Eclipse timing             & $\sigma$  & Ref.     \\    
      &BJD(TDB) 2400000+          &  (d)      &          \\
\hline       
-67     & 42520.76747  & $1.0\times10^{-5}$ & \footnotemark[1]  \\ 
-32     & 42532.81905  & $2.0\times10^{-5}$ & \footnotemark[1]  \\
-29     & 42533.85204  & $9.0\times10^{-5}$ & \footnotemark[1]  \\
0       & 42543.83769  & $1.0\times10^{-5}$ & \footnotemark[1]  \\ 
3       & 42544.87068  & $1.0\times10^{-5}$ & \footnotemark[1]  \\ 
851     & 42836.86314  & $6.0\times10^{-5}$ & \footnotemark[1]  \\ 
1966    & 43220.79202  & $1.2\times10^{-4}$ & \footnotemark[1]  \\ 
2132    & 43277.95101  & $6.0\times10^{-5}$ & \footnotemark[1]  \\
2896    & 43541.01972  & $1.2\times10^{-4}$ & \footnotemark[1]  \\  
28666   & 52414.425572 & $1.0\times10^{-6}$ & \footnotemark[2]  \\ 
29735   & 52782.515227 & $1.0\times10^{-6}$ & \footnotemark[2]  \\ 
29738   & 52783.548219 & $1.0\times10^{-6}$ & \footnotemark[2]  \\ 
30746   & 53130.633688 & $3.0\times10^{-6}$ & \footnotemark[2]  \\ 
32706   & 53805.522115 & $2.0\times10^{-6}$ & \footnotemark[2]  \\ 
32709   & 53806.555113 & $1.0\times10^{-6}$ & \footnotemark[2]  \\ 
34054   & 54269.680087 & $1.0\times10^{-6}$ & \footnotemark[2]  \\ 
37069   & 55307.837585 & $1.0\times10^{-6}$ & \footnotemark[3]  \\ 
38913   & 55942.783670 & $1.0\times10^{-6}$ & \footnotemark[4]  \\
37963   & 55615.669378 & $9.0\times10^{-6}$ & \footnotemark[4]  \\ 
38076   & 55654.578751 & $7.0\times10^{-6}$ & \footnotemark[4]  \\
38250   & 55714.492323 & $4.0\times10^{-6}$ & \footnotemark[4]  \\ 
39023   & 55980.660063 & $5.0\times10^{-6}$ & \footnotemark[4]  \\ 
40121   & 56358.735346 & $4.0\times10^{-6}$ & \footnotemark[4]  \\ 
40211   & 56389.725126 & $7.0\times10^{-6}$ & \footnotemark[4]  \\ 
40234   & 56397.644731 & $4.0\times10^{-6}$ & \footnotemark[4]  \\
40582   & 56517.47188  & $1.0\times10^{-5}$ & \footnotemark[5]  \\
41084   & 56690.325955 & $3.0\times10^{-6}$ & \footnotemark[4]  \\
41404   & 56800.511828 & $9.0\times10^{-6}$ & \footnotemark[4]  \\
42214   & 57079.419823 & $2.0\times10^{-6}$ & \footnotemark[4]  \\ 
44445   & 57847.621959 & $3.0\times10^{-6}$ & \footnotemark[5]  \\
44451   & 57849.68797  & $2.5\times10^{-5}$ & \footnotemark[5]  \\
46565   & 58577.60336  & $1.0\times10^{-5}$ & \footnotemark[5]  \\
46739   & 58637.51692  & $1.0\times10^{-5}$ & \footnotemark[5]  \\
46861   & 58679.525310 & $9.0\times10^{-6}$ & \footnotemark[5]  \\
46890   & 58689.510895 & $7.0\times10^{-6}$ & \footnotemark[5]  \\
47541   & 58913.670275 & $7.0\times10^{-6}$ & \footnotemark[5]  \\
47544   & 58914.703256 & $5.0\times10^{-6}$ & \footnotemark[5]  \\ 
47622   & 58941.561085 & $1.4\times10^{-5}$ & \footnotemark[5]  \\ 
\hline                                   
\end{tabular}\\
\footnotemark[1]\citet{Green+1978};~
\footnotemark[2]\citet{Parsons+2010};~
\footnotemark[3]\citet{Parsons+2012};~
\footnotemark[4]\citet{Bours+2015};~
\footnotemark[5]This study
\end{table}

\subsubsection{Light travel time effect}\label{lttsec}

The LTT effect is observed when additional components gravitationally interact with an object that has a stable clock, in our case, the primary eclipses of GK Vir, forcing it to rotate around the common mass centre of the entire system. Thus, the binary moves away from and closer to an external observer at rest. Because the speed of light is constant, the observer will see the period of the binary become larger or smaller when it is moving away or approaching, respectively. Adding the LTT mathematical relation ($\tau_j$) obtained by \citet{irwin1952} to Equation~\ref{lin_efem},
\begin{equation}
T_{\rm min} = T_0 + E\times P_{\rm bin} + \sum_1^n \tau_j,
\label{ephem_ltt}
\end{equation}
where,
\begin{equation}
\tau_{\rm j} =  \frac{a_{\rm bin;j}\sin i_j}{c}\left[\frac{1-e^2_j}{1+e_j \cos f_j}\sin (f_j+\omega_j) 
+ e_j\sin (\omega_j)\right].
\label{ltt}
\end{equation}
In the last equation, $a_{\rm bin;j}$ is the semimajor axis, $c$ is the speed of light, $i_j$ is the inclination, $e_j$ is the eccentricity, $\omega_j$ is the periastron argument, and $f_j$ is the true anomaly. These parameters are relative to the orbit of the centre of mass of the inner binary around the common mass centre consisting of the inner binary and of the $j-$th body. 

We fitted Equation~\ref{ephem_ltt} with one LTT effect to the mid-eclipse times. The resulting $\chi_{\rm red}^2$ was 2.1, and the residuals have no indication of another cyclic variation. To search for the best solution and to sample the parameters of Equation~\ref{ephem_ltt}, we used the \textsc{PIKAIA} algorithm \citep[][]{char1995} and an MCMC procedure \citep[][]{Foreman-Mackey2013}, respectively. The best solution is shown with red line in Figure~\ref{figo-c}, the posterior distributions of the fitted parameters are displayed in Figure~\ref{histo}, and the numerical values with their corres\-ponding standard uncertainties are presented in Table~\ref{parameters}. 

\begin{table}
\caption{Parameters of the linear ephemeris, LTT effect, and apsidal motion (Eqs.~\ref{ephem_ltt} and \ref{eqam}) adjusted to the mid-eclipse times of GK Vir.} 
\label{parameters}  
\begin{center}
\begin{tabular}{l c r }        
\hline             
  \small Linear ephemeris & LTT and apsidal motion    \\
\hline
Parameter  & Value & Unity  \\    
\hline                        
$P_{\rm bin}$ & $0.3443308426(3)$  & d  \\
$T_0$      & $2442543.83763(5)$ & BJD  \\
\hline
  & \small LTT $\tau_1$ term &  \\
\hline
Parameter  & Value & Unity  \\   
\hline                        
$P$           & $24.34^{+2.15}_{-1.64}$                & yr  \\
$T$           & $2453028^{+443}_{-370}$                & BJD   \\
$a_{\rm bin}\sin i$& $0.0109^{+0.0014}_{-0.0007}$   & au     \\
$e$           & $0.14\pm0.04$                 &        \\
$\omega$      & $198^{+22}_{-18}$                     & $\degree$ \\
$f(m)$        & $(1.6^{+1.3}_{-0.7}) \times 10^{-9}$ & M$_{\odot}$  \\ %
$(a_{\rm min})^{a}$   & $7.38^{+1.26}_{-0.72}$   & au  \\ 
$(m_{\rm min})^{b}$  & $0.95^{+0.22}_{-0.13}$     & M$_{\rm Jup}$ \\
$\chi_{\rm red}^2$ & 2.1        &  \\
\hline                                   

  & \small Apsidal motion &  \\
\hline
Parameter  & Value & Unity  \\
\hline
$P_{\rm AM}$ & $24.0\pm0.3$                & yr  \\
$e$          & $(5.35\pm0.02) \times 10^{-5}$                 &     \\
$\chi_{\rm red}^2$ & 3.4         &  \\
\hline
\end{tabular} \\
\footnotesize{$^{\rm a}$Minimum semimajor axis of the outer body} \\
\footnotesize{$^{\rm b}$Minimum mass of the outer body}
\end{center}
\end{table}

In the LTT scenario, one important test is to check if the third-body orbital solution is long-term stable. To do this, we used the solution shown in Table~\ref{parameters} and performed numerical integration using an $N$-body code with the usual Bulirch-Stoer integrator \citep[][]{Cincotta+2016}.

\begin{figure*}
\begin{minipage}{175mm}
\resizebox{\hsize}{!}{\includegraphics[angle=0]{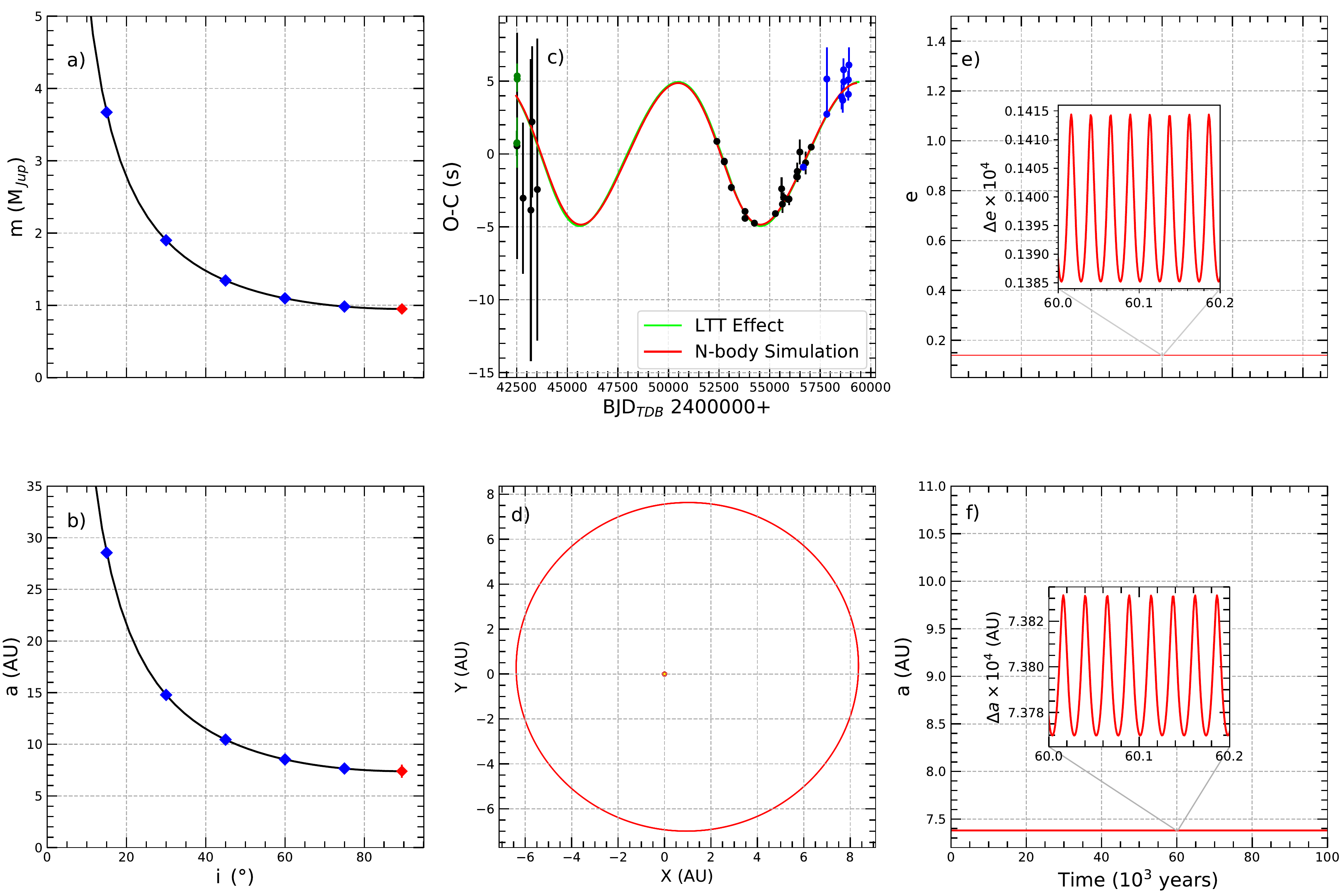}}
\caption{Numerical integrations of the outer body orbit around GK\,Vir: (a) the mass and (b) semimajor axis as a function of the inclination, (c) the O-C diagram, (d) the outer body orbit in the inner-binary mass centre reference system, and (e) the evolution of the eccentricity and (f) semimajor axis for $10^5$~yr. In panels from (c) to (f) are shown the case for inclination equal to 90$\degree$. In panels (e) and (f) two regions are zoomed in with their amplitudes of variation expanded by a factor of $10^4$ to better visualize the short-term effects.}
\label{nbody}
\end{minipage}{}
\end{figure*}

Initially, to verify if the LTT solution shown in Figure~\ref{figo-c} agrees with the numerical one, we performed an orbital integration with a time-step of 6~d and the total time of 45~yr, which is approximately the observational coverage of the GK\,Vir eclipse times. The O-C diagram obtained from the numerical simulation is presented with red line in panel (c) of Figure~\ref{nbody}, which agrees with the LTT analytical solution shown in this same panel with green line.   

In the second step, to test the third-body long-term stability, we performed numerical integrations considering the inclination between the external body and our line of sight equal to 15, 30, 45, 60, 75, and 90$\degree$. The time-step was set to 1.3~yr and the total integration time of $10^5$~yr. For each inclination, the mass of the outer body was obtained using a Newton-Raphson iteration in the following mass function:
\begin{equation}
 f(m_j) = \frac{4\pi^2(a_{\rm bin,j}\sin i_j)^3}{GP^2_j} = \frac{(m_j\sin i_j)^3}
 {(M_{\rm bin}+m_j)^2},
\label{mass_function}
\end{equation}
where $G$ is the gravitational constant, $M_{\rm bin}$ is the mass of the inner binary, $P_j$ and $m_j$ are the orbital period and mass of the external body, respectively. In panels (a) and (b) of Figure~\ref{nbody} are displayed the mass and the semimajor axis of the third body as a function of the inclination. In the same figure, panels (d), (e), and (f) show, respectively, the orbit of the third body in the inner-binary mass-centre reference system, the temporal evolution of the eccentricity and of the semi-major axis of the outer body for the inclination equal to 90$\degree$, which is approximately the same inclination of the inner binary. Inside of panels (e) and (f), two regions were zoomed in to better visualize the effect in the eccentricity and semimajor axis due to the orbital movement of the third body around the central binary. As one can see in Figure~\ref{nbody}, for the inclination equal to 90$\degree$, both eccentricity and semimajor axis of the third body are almost constant during $10^5$~yr, only varying with a small amplitude ($\sim$10$^{-7}$) in the same frequency of the outer body period, indicating that the third body has a stable orbit. The same result was found for the other inclinations.

\subsubsection{Apsidal motion}

The second possible scenario to explain the OPV of GK Vir is the apsidal motion. This effect consists of the rotation of the apsidal line due to tidal interactions in the close binaries and it can be observed when the orbital eccentricity of the binary is non-zero. One direct way to verify if the OPV of an eclipsing binary is due to the apsidal motion is by measuring the time between the primary and secondary eclipses throughout the cycle \citep[see e.g.,][]{Parsons+2014}. However, as the secondary eclipse of GK Vir was not measured yet, we adopted the following approaches to check if the apsidal motion is the one responsible for the OPV of this binary:   

(i) Check if the period generated by the apsidal motion is consistent with the period found in the O-C diagram.

The rotational variation rate of the apsidal motion has three contributions: tidal distortions generated by the non-spherical mass distribution of the stars ($\dot \omega_{\rm tide}$), rotation ($\dot \omega_{\rm rot}$), and effect due to the general relativity ($\dot \omega_{\rm GR}$). These three contributions, in degrees per year, can be calculated by  
\begin{equation}
\dot \omega_{\rm tide} = 15\frac{360}{P_{\rm bin}}\left(\frac{R_2}{a_{\rm bin}}\right)^5\frac{m_1}{m_2}\frac{1 + 1.5e^2 + 0.125e^4}{(1-e^2)^5} {\rm k_2},
\label{omegat}
\end{equation}
\begin{equation}
\dot \omega_{\rm rot} = \frac{360}{P_{\rm bin}}\left(\frac{R_2}{a_{\rm bin}}\right)^5\frac{m_1+m_2}{m_2}\frac{1}{(1-e^2)^2}{\rm k_2},
\label{omegar}
\end{equation}
and 
\begin{equation}
\dot \omega_{\rm GR} = \frac{360}{P_{\rm bin}}\left(\frac{3G}{c^2}\right)\frac{m_1+m_2}{a_{\rm bin}(1-e^2)},
\label{omegagr}
\end{equation}
where $R_2$ is the radius of the secondary star, $m_1$ and $m_2$ are the masses of the primary and secondary components, $G$ is the gravitational constant, $c$ is the speed of light, $k_2$ is the apsidal constant, and the other parameters are the same as defined in the previous sections. The apsidal constant, which is related to the concentration of mass of the tidally distorted star, has been the subject of several theoretical studies \citep[e.g.,][]{Sirotkin+2009, Claret+2010}. Following the work done by \citet[][]{Feiden+2011}, and extrapolating the values of mass presented in its Table 1 to the secondary mass of GK Vir ($m_2 = 0.116$~\rm M$_\odot$), we obtained $k_2 \sim 0.156$. Using the parameters derived by \citet[][see section~\ref{introduction}]{Parsons+2012}, $k_2 = 0.156$, and $e = 0$ in Eqs.~\ref{omegat}, \ref{omegar}, and \ref{omegagr}, the upper limit for the period of the apsidal motion would be $\sim$16.4 yr, which is smaller than the period obtained in the O-C diagram of GK\,Vir. For this calculation, we only consider the $\dot \omega_{\rm tide}$ and $\dot \omega_{\rm rot}$ generated by $m_2$, as the WD contribution is much smaller than the secondary star one.

(ii) Verify if the circularization time for GK Vir's orbit is smaller or higher than the WD cooling time.

Theoretically, it is possible to check if the GK Vir system had enough time to completely circularize its orbit. To do so, we need to compute the cooling time of the WD, which can be considered as the age that the system has in the current configuration, and then, compare it with the circularization time of its orbit. 

Following \citet[][]{Althaus+2010}, the WD cooling time can be estimated using the Mastel's law \citep[][]{Mestel1952} through the approximation
\begin{equation}
\tau_{\rm cool} \approx \left(\frac{10^8}{A}\right)\left(\frac{m_1}{M_{\rm \odot}}\right)^{5/7}\left(\frac{L_1}{L_{\rm \odot}}\right)^{-5/7}, 
\end{equation}
where $A$ is the mean atomic number, and $L_1$ is the luminosity of the WD. Replacing the values obtained by \citet[][]{Parsons+2012} for the WD of GK Vir and assuming for $A$ the atomic number of Carbon, yields $\tau_{\rm cool} \sim 5.7$ Myr.

The orbital circularization of a binary system is an effect caused due to the interaction of the tides between its components \citep[see e.g.,][]{Zahn1984}. The time to circularize the binary's orbit can be estimated by
\begin{equation}
\tau_{\rm cir} = \left[21k_2q(q+1) \left(\frac{L_2}{m_2R_2^2}\right)^{1/3}\left(\frac{R_2}{a_{\rm bin}}\right)^{6}\right]^{-1}, 
\end{equation}
where $k_2$ is the apsidal constant, $q = m_2/m_1$ is the mass ratio, and $L_2$, $m_2$, and $R_2$ are the luminosity, mass, and radius of the secondary star, respectively. Using the parameters derived by \citet[][]{Parsons+2012} for GK Vir, we obtained the circularization time $\sim$0.7 Myr. Thus, as $\tau_{\rm cool}$ is $\sim$8.1 times larger than $\tau_{\rm cir}$ and considering only the tidal interaction acting on the orbital parameters of the binary, we conclude that GK Vir had enough time to circularize its orbit.

(iii) Verify if the equation of the apsidal motion fits well to the mid-eclipse times of GK Vir.

Following the study done by \citet[][]{Todoran1972}, the equation that describes the linear ephemeris plus the apsidal motion using the first-order approximation for the orbital eccentricity ($e$) is 
\begin{equation}
T_{\rm min} = T_0 + E\times P_{\rm bin} + \frac{eP_{\rm bin}}{2\pi}({\rm cosec}^2i_{\rm bin}+1)\sin\left(\frac{2\pi t}{P_{\rm am}}\right),
\label{eqam}
\end{equation}
where, $i_{\rm bin}$ is the binary orbital inclination, $t$ is mid-eclipse time obtained from the linear ephemeris, and $P_{\rm am}$ is the period of the apsidal motion. We fit this equation to the mid-eclipse times of GK Vir adopting the same procedure used for the LTT analysis. The fitted parameters with their uncertainties are presented in Table~\ref{timing} and the best solution is shown with the black line in Figure~\ref{figo-c}. The best solution provides $\chi^2_{\rm red} = 3.4$, which is larger than the one obtained for the LTT best solution (see Section~\ref{lttsec}).

\subsubsection{Applegate mechanism}

The third possible scenario for the OPV of GK Vir is associated with the magnetic cycle of active stars. This effect proposed by \citet{Applegate1992}, called Applegate mechanism, consists of the OPV of the system due to the changes in the form of a magnetically active component. The shape of the star may change due to the variation of the quadrupole moment, which in turn leads to changes in the orbital period of the binary. These changes must occur at the same time-scale as the magnetic activity cycle (MAC) of the star.

In a series of papers \citep{1998MNRAS.296.893L,1999AA.349.887L,2006MNRAS.369.1773L}, Lanza and collaborators refined the treatment done by \citet{Applegate1992}. \citet{2006MNRAS.369.1773L} discarded the Applegate mechanism for RS~CVn systems. Moreover, the author commented that the Applegate hypothesis cannot explain the orbital period modulation of close binary systems composed by a late-type secondary star.

In the same direction, \citet[][]{Brinkworth2006} included a stellar thick outer shell to the Applegate theory. More recently, \citet{Volschow+2016} developed a new formulation to add the quadrupole moment changes in two finite regions, core and external shell. With this new model, the authors using 16 compact binaries concluded that the Applegate mechanism can explain the eclipse time variation for four systems.

In this context, the way to verify if this mechanism can explain the modulation in the binary orbital period is checking if the observed variation amplitude in the O-C diagram can be produced by the energy of the secondary star. Following the work done by \citet{Volschow+2016} and using their online calculator\footnote{http://theory-starformation-group.cl/applegate/index.php}, we obtained that the required energy for the finite-shell constant density model is $\sim$10$^5$ times larger than the energy of the secondary star.

Recently, a new model based on the exchange of angular momentum between the active component spin and the orbital motion was proposed by \citet[][]{Lanza2020}. This author found that the systems with energy $\sim$10$^2$ to 10$^3$ times smaller than the required energy to explain the OPV, reported by previous models, can be explained using this new approach.

\section{Discussion and conclusions}\label{discussion}

We present 10 new mid-eclipse times of GK\,Vir from 2013 August to 2020 April. We combined these measurements with all mid-eclipse times available in the literature and performed an orbital period analysis. One cyclic modu\-lation is seen in the O-C diagram (see Figure~\ref{figo-c}). Based on the modulation period, which is $\sim$24~yr, we investigate if this variation could be explained by the Applegate mecha\-nism, the apsidal motion, or the LTT effect.

For the Applegate mechanism, following \citet{Volschow+2016} we showed that the amount of required energy to explain the O-C diagram of GK\,Vir is  $\sim$10$^5$ times larger than the energy of the secondary star. Based on this amount of required energy, even considering the new model proposed by \citet[][]{Lanza2020}, this mechanism would hardly explain the O-C diagram of GK Vir.

\begin{figure}
\resizebox{\hsize}{!}{\includegraphics[angle=0]{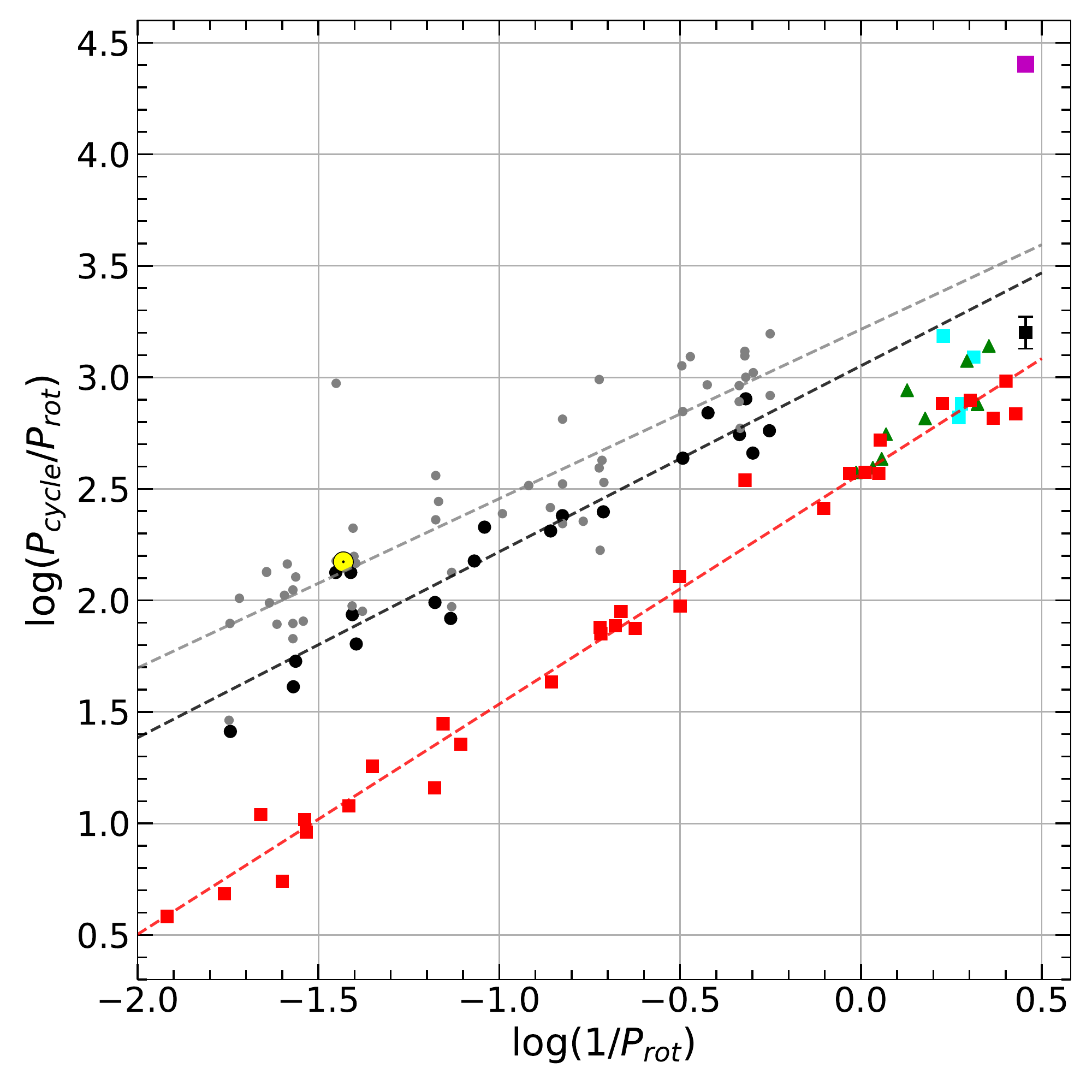}}
\caption{MAC versus rotational period diagram as shown in \citet[][]{Vida+2014} and \citet[][]{Almeida+2019}. The black dots, blue squares,  green triangles, and red squares represent measurements from \citet[][]{Vida+2013,Vida+2014,Olah+2009}, and \citet[][]{Savanov2012}, respectively. The gray dots show the data from different surveys presented in \citet[][]{Olah+2009}. The gray line represents the fit (using linear regression) to all the data from \citet[][]{Olah+2009} and \citet[][]{Vida+2013,Vida+2014} excluding M stars, while the black and red lines show the fit to the shortest cycles of that data set. The Sun is shown with its standard symbol and the measurement derived by \citet[][]{Almeida+2019} is presented with a black square. The magenta point (see top right) is the hypothetical measurement for GK\,Vir (see Section~\ref{discussion} for more details). As this point does not agree with the trends of the experimental measurements, this hypothetical scenario can be ruled out.}
\label{figorbit}
\end{figure}

Besides the energy test, we can verify if the period found in O-C diagram could be explained by one hypothetical MAC of the secondary star, which is directly correlated to the Applegate mechanism. To do so, we consider the period found in the O-C diagram ($\sim$24~yr) as the MAC of the secondary star. Furthermore, we assume that the secondary star is synchronized with the orbital period of GK\,Vir and therefore its rotational period would be $\sim$8.3~h. Adding these values in the MAC versus rotation period diagram (see the magenta point at the top right hand corner in Figure~\ref{figorbit}), we conclude that it does not agree with the empirical trends. Also, the first evolved system, similar to GK\,Vir, with measured MAC reported by \citet[][see the black square in Figure~\ref{figorbit}]{Almeida+2019} has the same trend than the other measures for single stars in this diagram. Therefore, it is one additional evidence against the possibility of the O-C diagram of GK Vir being explained by the Applegate mechanism. However, we emphasize that this is a particular case, and thus, this result does not rule out this mechanism as a possible cause of the orbital variation of other eclipsing post-common envelope binaries, as for example, it is the most likely cause of V471 Tau \citep[][]{Hardy+2015}.

In the apsidal motion context, we used three approaches to analyse if the OPV of GK Vir could be explained by this effect. In the first two cases, we showed that the predicted theoretical period by the apsidal motion is $\sim$1.5 times smaller than the period found in the O-C diagram, and the cooling time of the WD is $\sim$8.1 times larger than the circularization time, which are pieces of evidence against the explanation via the apsidal motion. In the third analysis, we fitted the equation of the apsidal motion to the mid-eclipse times of GK Vir and obtained $\chi^2_{\rm red} = 3.4$ (see Figure~\ref{figo-c}). Despite this relatively low $\chi^2_{\rm red}$, which would indicate a good fit, it is larger than the one obtained for the LTT effect (see below).

Finally in the LTT scenario, we showed that the equation that represents a circumbinary body fits well to the mid-eclipse times of GK Vir, see Figure~\ref{figo-c}. The best solution, which provides $\chi^2_{\rm red} = 2.1$, yields orbital period $P = 24.34^{+2.15}_{-1.64}$~yr, and eccentricity $e = 0.14\pm0.04$, for the outer body. Adopting the mass and the inclination of the inner binary, $0.68~\rm M_{\odot}$ and 89$\fdg$5 \citep{Parsons+2012}, and under the assumption of coplanarity between the outer body and the inner binary, the mass of the circumbinary body is $m_3\sim0.95~ M_{\rm Jup}$. Therefore, in this scenario, GK\,Vir would be composed of an inner binary and a Jupiter-like planet. However, as the observational baseline of GK\,Vir is smaller than twice the period found in the O-C diagram and the first eclipse time measurements from \citet[][]{Green+1978} have large error bars, we must take this solution as preliminary.

As some studies have suggested additional circumbinary bodies as a possible explanation for the OPVs of post-common envelope binaries \citep[e.g.,][]{2009AJ.137.3181L, Almeida+2011IAUS} and further works have shown that their orbits are in an unstable configuration \citep[e.g.,][]{Horner+2012,Horner+2013MNRAS}, an important test for the LTT scenario is to verify if the third-body orbital solution shows long-term stability. To do that, we performed a dynamical analysis by using $N$-body numerical integrations. The results for six different inclinations (15, 30, 45, 60, 75, and 90$\degree$) between the third-body orbital plane and our line of sight showed that the outer body around GK\,Vir has a stable orbital configuration over, at least, $10^5$ yr (see Figure~\ref{nbody}). Therefore, this reinforces the LTT effect as the most likely explanation for the OPV behavior of GK\,Vir. 

If the LTT effect is confirmed with future data as the true cause of the orbital modulation of GK\,Vir, the third body would be the planet with one of the longest orbital periods, with a full observational baseline, discovered so far. Considering the possible formation scenarios, according to \citet{Perets2011} this circumbinary body could have been formed either at the same time as the inner binary formation (called as the first generation of planets) or after the common envelope phase of the inner binary (known as the second generation of planets). Although our data are not conclusive on these two possible formation scenarios, our results place GK\,Vir as a promising target for further study on this subject with the new generation of large telescopes, e.g., Giant Magellan Telescope, Thirty Meter Telescope, and Extremely Large Telescope.

\section*{Acknowledgements}
We thank the anonymous referee for helpful suggestions which were important to improve the paper. This study was partially supported by Coordena\c{c}\~ao de Aperfei\c{c}oamento de Pessoal de N\'ivel Superior (CAPES). LAA, AD and TAM thank Funda\c{c}\~ao de Amparo \`a Pesquisa do Estado de S\~ao Paulo (FAPESP) through the projects (LAA and AD: 2011/51680-6, LAA: 2012/09716-6, 2013/18245-0, and TAM: 2016/13750-6). GMV thanks Conselho Nacional de Desenvolvimento Cient\'ifico e Tecnol\'ogico (CNPq) for funding. This study was based on observations carried out at the Observatorio do Pico dos Dias (OPD/LNA) in Brazil. We thank the OPD staff for support and help during the observations.
 
\section*{Data availability}
The data underlying this article will be shared on reasonable request to the corresponding author.

\appendix

\section{The MCMC posterior distributions}

\begin{figure*}
\begin{minipage}{175mm}
\resizebox{\hsize}{!}{\includegraphics[angle=0]{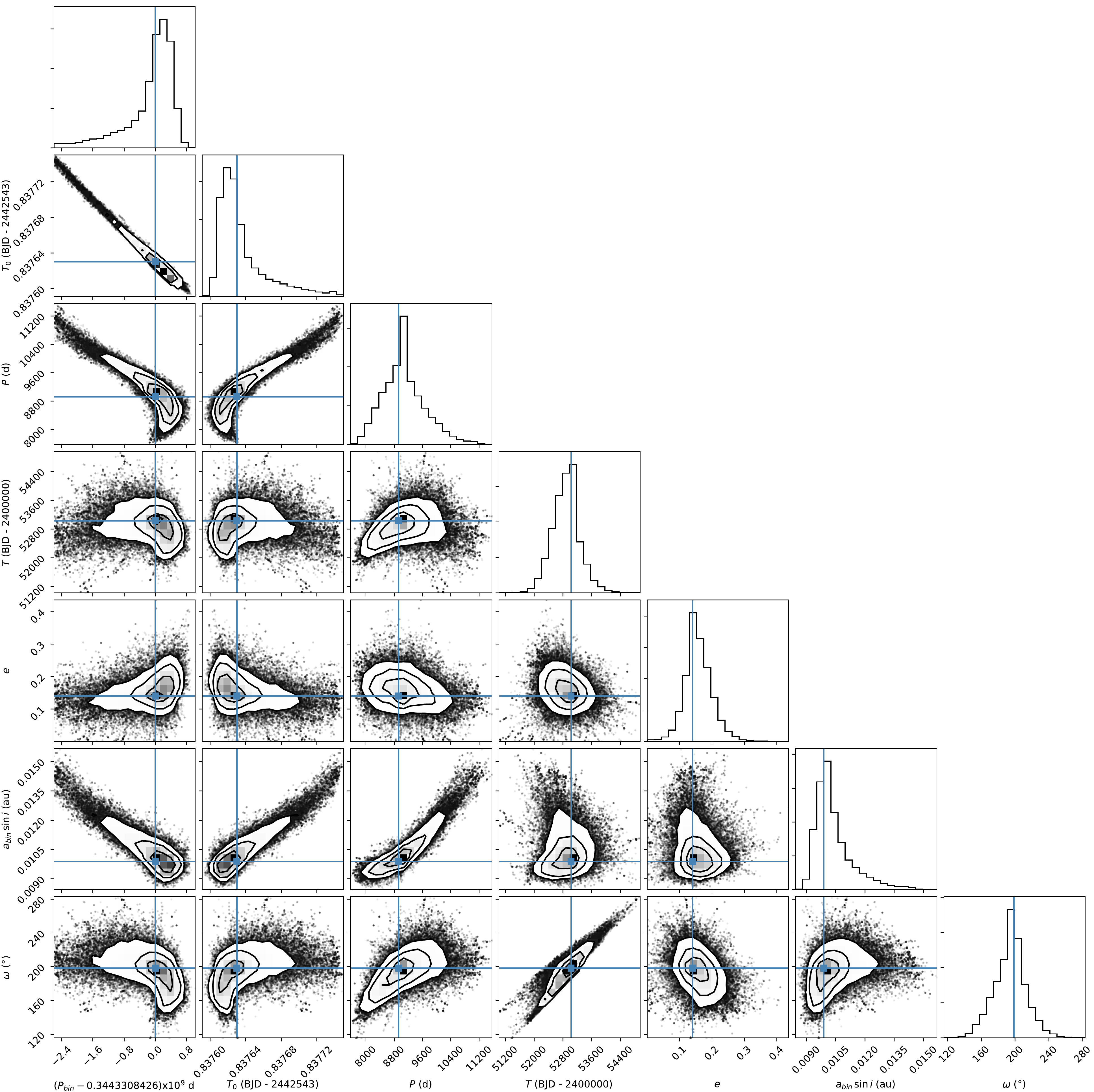}}
\caption{Set of distributions of the a posteriori probability densities for the free parameters of Eq.~\ref{ephem_ltt} fitted to the mid-eclipse times of GK Vir. The blue points at the centre of the crosses represent the values of the parameters for the best solution.}
\label{histo}
\end{minipage}{}
\end{figure*}


\end{document}